\documentclass[twocolumn,showpacs,preprintnumbers,amsmath,amssymb]{revtex4}
\usepackage{graphicx}
\usepackage{dcolumn}
\usepackage{bm}

\begin{document}


\title{Non-adiabatic geometrical quantum gates in semiconductor quantum dots}

\author{Paolo Solinas,$^{*}$ Paolo Zanardi,$^{\dag}$ Nino Zangh\`{\i},$^{*}$
and Fausto Rossi$^{\dag,\ddag}$
}
\affiliation{
$^*$ Istituto Nazionale di Fisica Nucleare (INFN) and
Dipartimento di Fisica, Universit\`a di Genova,
Via Dodecaneso 33, 16146 Genova, Italy \\
$^\dag$ Institute for Scientific Interchange (ISI),
Viale Settimio Severo 65, 10133 Torino, Italy \\
$^\ddag$ Istituto Nazionale per la Fisica della Materia (INFM) and
Dipartimento di Fisica, Politecnico di Torino, Corso Duca degli
Abruzzi 24, 10129 Torino, Italy }

\date{\today}

\begin{abstract}
In this paper we study the implementation of  non-adiabatic geometrical 
quantum gates with  in semiconductor quantum dots.
Different quantum information enconding/manipulation schemes
exploiting excitonic degrees of freedom are discussed. 
By means  of the  Aharanov-Anandan geometrical phase one  can 
avoid the limitations of adiabatic schemes  relying on adiabatic Berry phase;
fast geometrical quantum gates can be in principle implemented.
\end{abstract}

\maketitle
\section{Introduction}
The  Holonomic Quantum Computation proposal (HQC) \cite{HQC} recently led
to  a number of investigations  \cite{ad_prop}   aimed to assess
its  feasibility. 
At variance with  ``ordinary'' dynamical quantum gates the Holonomic ones depend only on 
geometrical features (i.e. the angle swept by a vector on a sphere) 
 of a suitable quantum control process.
It  has been argued that HQC might lead to computational schemes
more robust against some class of errors. Despite this crucial property
has not been clearly demonstrated so far (for a critical view see e.g., \cite{critic}),
 HQC surely provides a sort of an intermediate step towards topological quantum computing \cite{top,zl}
The latter represents an intriguing and ambitious   paradigm for 
inherently fault-tolerant QC.

Many proposals for practical HQC follow the adiabatic approach
\cite{ad_prop}; it
consists in changing the Hamiltonian parameters in order to
produce a loop in the Hamiltonian space ($H(0)=H(T)$).
For an  adiabatic evolution if we start from an eigenstate
 $|n(0)\rangle$ of $H(0)$ with eigenvalue $E_n(0)$, during the evolution we remain in
 the instantaneous eigenvector $|n(t)\rangle$ of $H(t)$ with  eigenvalue $E_n(t)$.
At the end of the loop the state will differs by the initial state
only for a phase factor (Berry phase).
If the eigenstate is degenerate we end in a superposition 
of the degenerate states and then we have a non-Abelian holonomic operator \cite{wize}.

On the other hand it is  well-known that the major obstacle against the practical
realization of quantum information processing (QIP)\cite{QC}  is provided by the
detrimental interaction with environmental degrees of freedom.
This interaction results, typically in a extremely short time,
in the destruction of the quantum coherence of the information-encoding quantum state,
that in turns spoils the computation \cite{QC}.  
It follows that for QIP purposes it  is very important to have fast logical
gates to be able to realize numerous logical operations within
decoherence time.

The fact that we have to change parameters slowly 
it is therefore an  obvious  drawback of the adiabatic approach. 
Then the possibility of have geometrical gate without
the adiabatic limitation looks very appealing.

In 1987 Aharanov and Anandan (A-A) \cite{AA} showed that there is a additional
geometrical phase factor for {\it all} the cyclic evolution of the
states (not only for the adiabatic ones).
The A-A phase is a generalization of Berry phase and we recover
this when the adiabatic condition is restored.
Recently some proposal for non-adiabatic geometrical gates have
been made \cite{non_ad_prop}.

In this paper we shall propose a universal set of non-adiabatic geometrical
gates using excitonic states in semiconductor quantum dots.
The schemes here below illustrated rely on 
on the physical setup  analyzed in Refs \cite{sol}
and on  the abstract geometrical structure
of Ref. \cite{xq}.
 
\section{exciton- no exciton qubit}

In Ref. \cite{biolatti} it has been  shown how excitonic states in a quantum dot
can be used to perform universal QIP.
The logical states were the ground state $|G\rangle$ and the 
excitonic state $|E\rangle$ and they were driven by {\it all-optical} 
control (with ultra-fast laser). Even if the decoherence time in this system is
quite short  the ultra-fast laser technology used for the coherent manipulations allows, in principle,
to perform a large number of operations.

Let us start by showing how the scheme by Qi et al. \cite{xq} can be
applied in  this semiconductor context. 
We have a two-level system ($\hbar=1$ and $\omega_0$ energy separation) 
interacting with a laser field (radiation-matter interaction) and then the 
interaction Hamiltonian can be written

\begin{equation}
  H_{int}
   = -[\Omega e^{-i\omega_L t- \phi}  |E \rangle \langle G |+ h.c.] 
\label{basicH}
\end{equation}

In a rotating frame (with precession frequency $\omega_L$) the total 
Hamiltonian is (using 'spin' formalism)
$ H_R = {\bf B} \cdot \vec{\sigma},$
with ${\bf B} = (\Omega~ \cos\phi, \Omega~ \sin \phi,
\frac{\omega_0-\omega_L}{2})$ and $\vec{\sigma} =
(\sigma_x,\sigma_y, \sigma_z)$.
This is the Hamiltonian presented in Ref. \cite{xq} and then we can 
obtain the same gates.
With ${\bf B}\neq 0$ the 'spin' will precede on the Bloch sphere
on a plane orthogonal to ${\bf B}$ according  the Bloch's equations.

Following Ref. \cite{xq} is easy to see that 
-- by  choosing the laser parameters
(phase and frequencies) in a suitable way --
one can produce a sequence of laser pulse that 
enact  a loop on the Bloch sphere;  the final state will acquire a geometrical
phase independently of the velocity during the traversed loop
(no adiabatic approximation).
The final operator depends on the angle
swept on the sphere by the state vector during the evolution.
With a sequence of two $\pi-$pulses we can obtain two single qubit gates.
First we take $\omega_L \neq \omega_0$ (off-resonant laser) and then 
produce two $\pi-$pulses with different phase (i.e. $\Delta \phi = \pi$) 
and obtain the following gate :

\begin{figure}[t]
  \includegraphics[height=3.5cm]{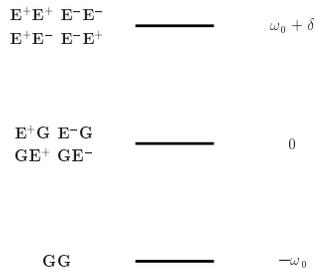}
  \caption{\label{fig:levels} Energy levels for two coupled dots with 
    dipole-dipole interaction. $\delta$ is the biexcitonic shift.}
\end{figure}

\begin{equation}
  \begin{array}{ccc}
 |0\rangle &\rightarrow & \cos\gamma |0\rangle - \sin\gamma |1\rangle   \\
 |1\rangle &\rightarrow & \cos\gamma |1\rangle + \sin\gamma |0\rangle
\label{eq:gate1_2}
 \end{array}
\end{equation}

where $\gamma$ is half the angle swept by 
the vector on the Bloch sphere and it depends on the gate 
parameters (i.e. the laser frequency) $\gamma = 2 \arctan (2 \Omega /
(\omega_0-\omega_L))$.

For a {\it selective phase} gate we have resonant condition ($\omega_0 = \omega_L$)
and produce two $\pi-$pulses with opposite phase 
($\phi_1 = - \phi_2 = \phi_0$) and we have

\begin{equation}
  \begin{array}{lcll}
    |0\rangle &\rightarrow &e^{i\tilde{\gamma}} &|0\rangle  \\
    |1\rangle &\rightarrow &e^{-i\tilde{\gamma}}& |1\rangle
    \label{eq:gate2}
  \end{array}
\end{equation}

where $\tilde{\gamma} = 2 \phi_0$.

We note that the dynamical phase factor in standard geometric quantum computation
must be eliminate with several adiabatic loop in order to let
the phase factor cancel each other. In this model it does not
appear because the motion on the Bloch sphere is on a plane
orthogonal to {\bf B} and so it can be easily shown that 
$\langle \psi| H |\psi \rangle =0 $ and the dynamical phase factor is zero.
Of course this geometric gates are much faster that the adiabatic one 
\cite{sol} which had the limitation of the slow change of parameters.

This kind of geometrical manipulation of excitonic-encoded  information
should be  easier to implement and to verify experimentally, because 
they are just produced by a sequence of $\pi-$pulses with constant parameters 
(frequency or phase of the laser) with just one laser instead of 
three lasers in which change the intensity and the phase during the evolution.

For the two-qubit gate we have to exploit qubit-qubit interaction in order 
to construct non-trivial operators; then every system has different 
implementation of such gates.
Since we work with semiconductor excitons we use exciton-exciton 
dipole interaction.

\begin{figure}[h]
  \includegraphics[height=3.8cm]{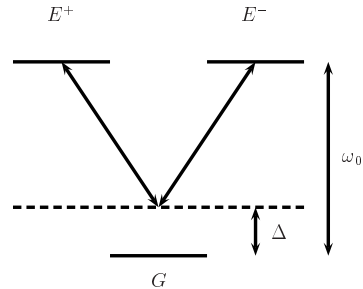}
  \caption{\label{fig:subspaces} Connection of the logical subspaces $E^+$
    and $E^-$. The $\Delta$ is the detuning of the lasers that allow us to
    connect the to states through Raman transition.}
\end{figure}

Let us consider two dots with exciton energy $\omega_0/2$ 
(the energy  is rescaled in order to have $-\omega_0/2$ 
for the ground states).
If the two dots are coupled the presence of an exciton in one of them
causes a energetic shift ($\delta$) in the other because of the dipole-dipole 
interaction. States with a single exciton are not shifted.
The energy levels are shown in figure \ref{fig:levels}.
The Hamiltonian accounting for the biexcitonic shift is $H_0 = 
(\omega_0+\delta) 
|EE\rangle \langle EE| - \omega_0 |GG\rangle \langle GG|$.

The dipole interaction between dots can be used to construct non-trivial
two-qubit gates both dynamical \cite{biolatti} and geometrical
\cite{sol}. In fact, if we use two lasers tuned to the two-exciton
state transition ($\omega_L^1 = \omega_L^2 = (\omega_0 + \delta)/2$), 
we can avoid single photon processes (which product $|EG\rangle$ and 
$|GE\rangle$ states) and favour only two-photon processes (which 
product $|EE\rangle$). 

The effective interaction Hamiltonian for the two-photon process is :

\begin{eqnarray}
   H_{int} = -\frac{2 \hbar^2}{\delta} 
   \Omega_{+}^2 e^{-i(\omega_{L,1}+\omega_{L,2})} 
   e^{-i(\phi_1 + \phi_2)} |E\rangle \langle G|^{\otimes\,2}  +
   \mbox{h.c.} 
   \label{eq:two_ph}
\end{eqnarray}

where $\omega_{L,i}$ e $\phi_i$ are the frequency and the phase of the 
laser $i$.

The total Hamiltonian is similar to the one in 
(\ref{basicH}) and then using a properly chosen sequence of 
sincronous pulses 
(so that the two-photon Rabi frequencies in \ref{eq:two_ph} simulate the 
one in \ref{basicH}) we can apply a phase gate similar to 
(\ref{eq:gate2}) and complete the universal set of 
quantum gates.

\section{ exciton spin qubit}
A further excitonic  encoding can be obtained following
 the spin-based scheme presented in \cite{sol}.
There a four-level system with three degenerate excited states 
($|E^{\pm}\rangle$ and $|E^0 \rangle $) and a ground state ($|G\rangle $)
was used; the excitonic states were connected with $|G\rangle $ by 
three different lasers with circular ($\pm$) and linear (along $z$ axis) 
polarization and, modulating the phase and the frequency of the three 
laser, we were able to construct adiabatic holonomic gates.

To obtain non-adiabatic geometrical gates in this 
system the basic idea is to encode logical information in two degenerate 
exciton states  with different total angular momentum i.e. $|E^{\pm}\rangle.$
The  extension  of the previous gating model is not completely straightforward;
in fact the logical qubits
$|E^+ \rangle$ and $|E^- \rangle,$ due to angular-momentum conservation in 
radiation-matter interaction,
are not directly 
 i.e., by a one-photon ladder operators, connected.

\begin{figure}[t]
  \includegraphics[height=4.5cm]{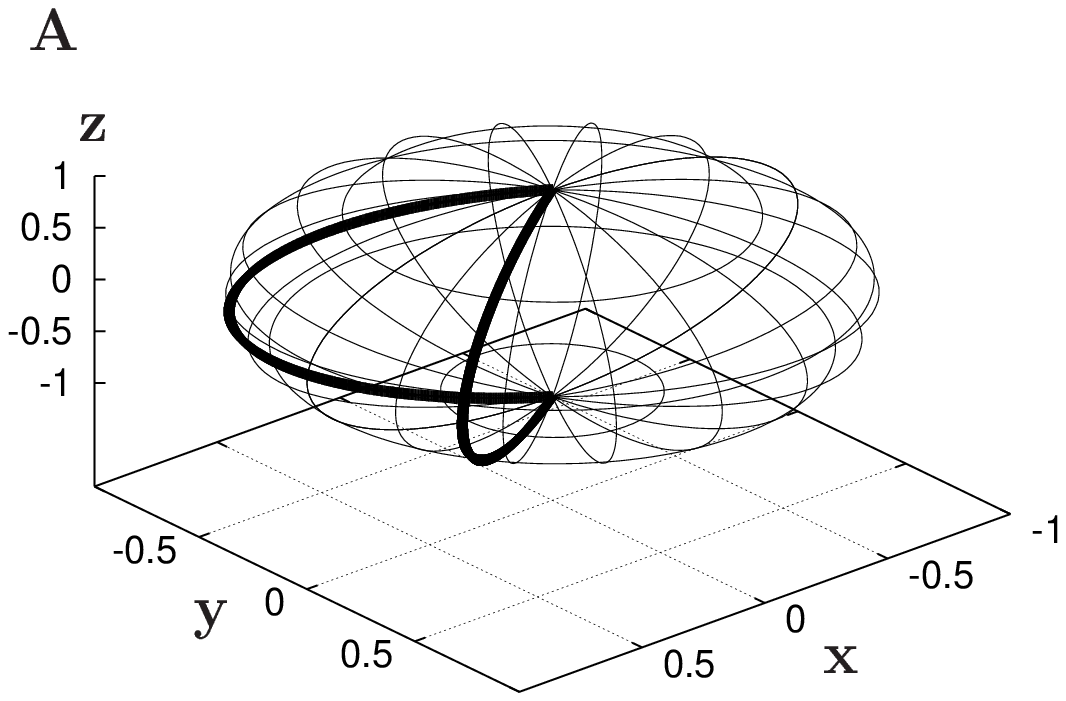}\\
  \includegraphics[height=4.0cm]{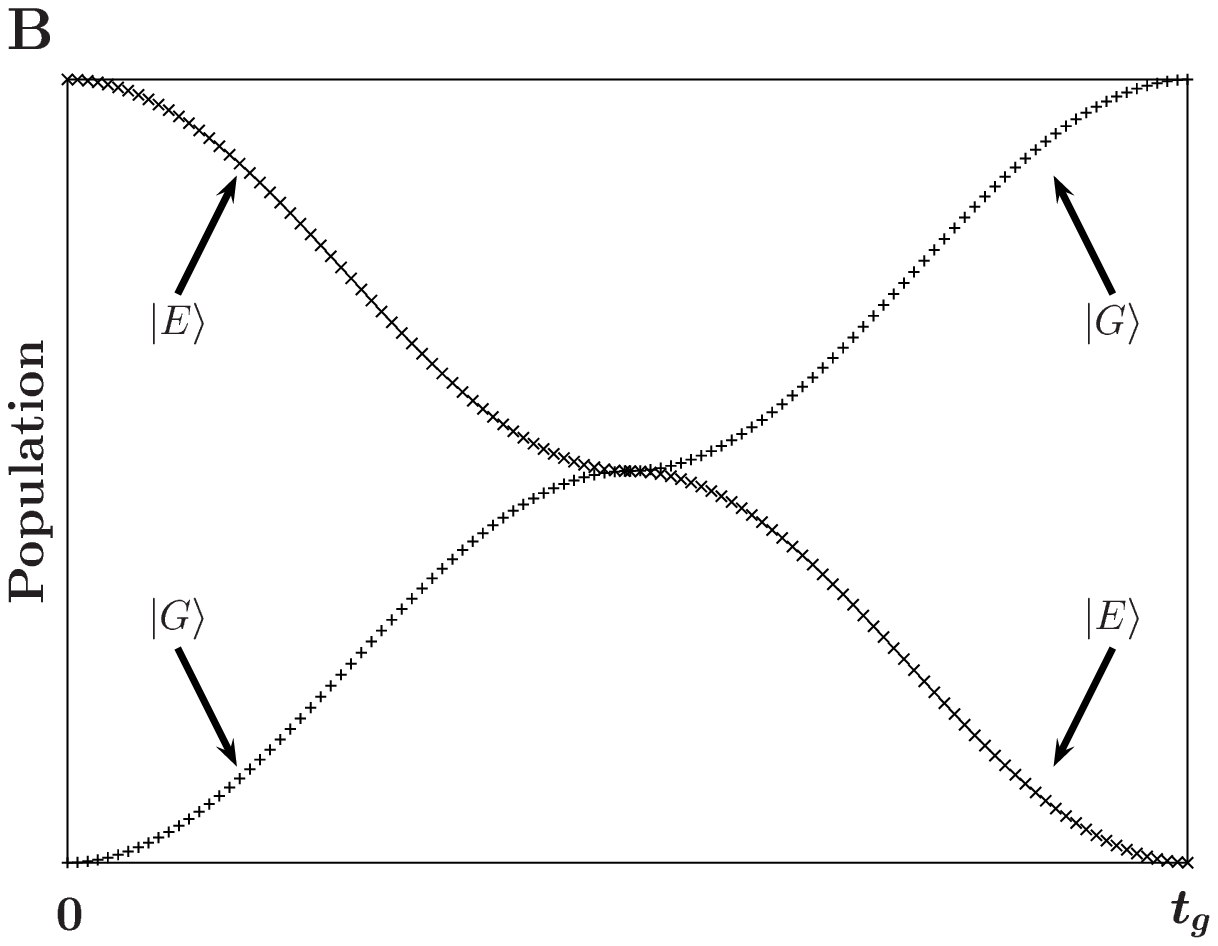}\\
  \caption{\label{fig:fig_gate1} Gate 1 for the unpolarized excitons 
    model. The parameters are chosen in order to obtain a $NOT$ gate.
    (A) Evolution of $|E\rangle$ state on the 
    Bloch sphere. (B) Population evolution for the logical 
    states $|E\rangle$ and $|G\rangle$.}
\end{figure}

In order to circumvent this problem and to enact such a ladder operator one can resort 
to an off-resonant two-photon Raman process. This is a standard trick in quantum optics.
Each quantum dot is shined by a couple of lasers
having polarizations $+$ and $-$ and a frequency with a detuning $\Delta$ with respect 
the  excitonic transition energy 
The level scheme with the associate transition is shown in Fig. \ref{fig:subspaces}. 
Provided that $\Omega_\pm\ll \Delta$ (the $\Omega_\pm$'s are the laser Rabi frequencies)
first order processes are then strongly suppressed; the dynamics is well-described by the
following second-order effective Hamiltonian
\begin{equation}
H_{eff} = \frac{\Omega_+\Omega_-}{\Delta}|E^+\rangle\langle E^-|+\mbox{h.c.}.
\label{H2}
\end{equation}
It should be now clear --since the above Hamiltonian as the same structure of  (\ref{basicH}) --
that even for this kind of excitonic encoding using different polarizations  one can realize all the required single-qubit operations.
    
Another single qubit gate that can be implemented easily is the
{\it phase shift} gate. Our scheme has {\it a priori}
separated sub-spaces because the different renspose to polarized laser.
So if we want $|E^+\rangle$ to get a phase factor, we can just 
switch the $+$ laser to resonant frequency, and then apply the
pulse sequence that produce gate 2. Since we can neglect the phase accumulated
  by $|G\rangle$ and no phase is accumulated by $|E^-\rangle$
the gate operator will be $U=exp(i\tilde{\gamma} |E^+\rangle \langle E^+|)$
where, as before,  $\tilde{\gamma}$ is half the solid angle swept in the 
evolution. These two gates complete the single-qubit gate set.

Finally, to obtain a universal set of quantum logical gates we must construct 
a two qubit gates. The easiest to be implemented in our model is a 
{\it selective phase gate}.
As shown before using lasers resonant with the two exciton with positive 
polarization we can select 
two-photon processes and couple only the $|E^+E^+ \rangle - |GG\rangle$
states \cite{sol}.
The effective Hamiltonian for these two-photon processes is similar 
to (\ref{eq:two_ph}) with $|E^+\rangle$ instead of a generic exciton 
state $|E \rangle$.

The two lasers are polarized with $+$ polarization and follow the 
pulse sequence for gate $1$ ; the final geometric 
operator will be $U=exp(i\tilde{\gamma} |E^+E^+ \rangle \langle E^+E^+|)$, 
where $\tilde{\gamma}$ is half the angle swept on the Bloch sphere in the
$|E^+E^+ \rangle - |GG\rangle$ space.

\begin{figure}[t]
  \includegraphics[height=4.5cm]{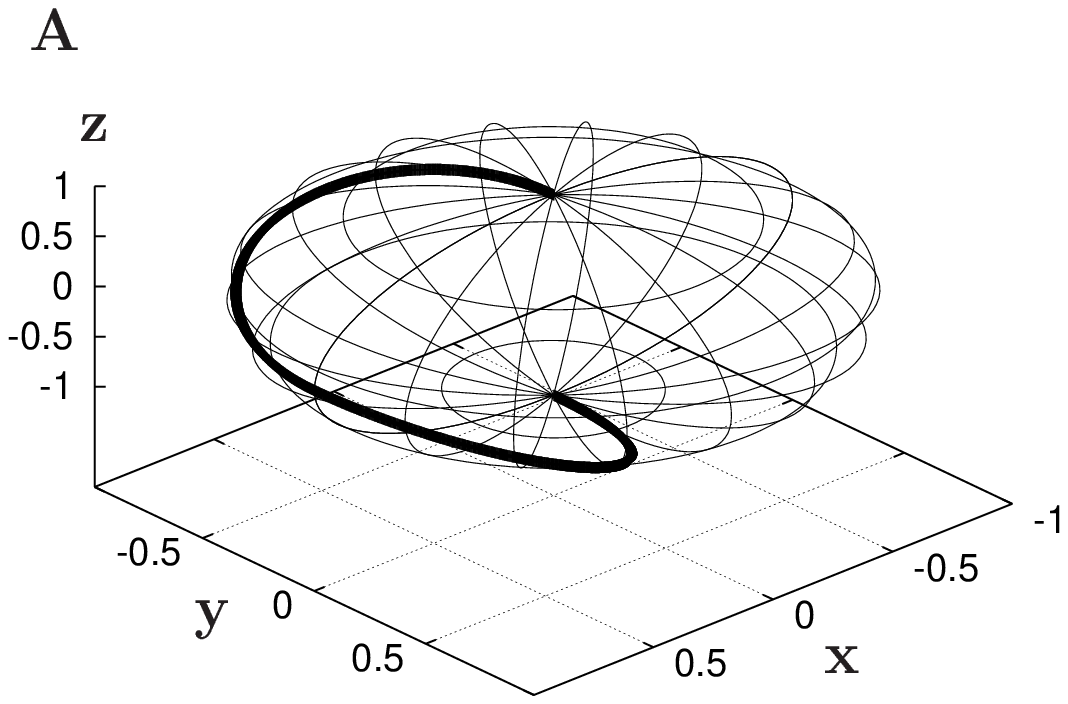}
  \includegraphics[height=4.0cm]{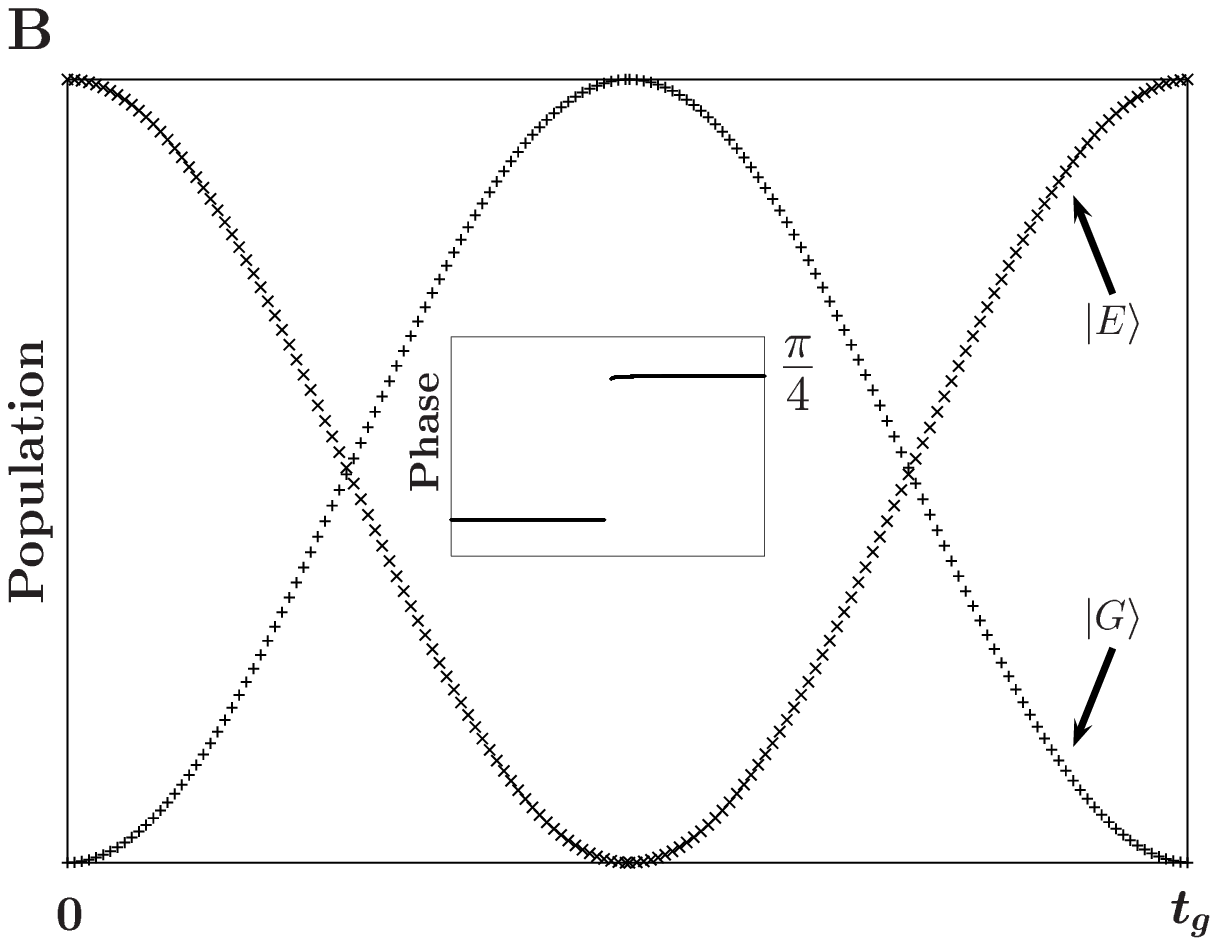}
  \caption{\label{fig:fig_gate2} Gate 2 for the unpolarized excitons 
    model.(A) Evolution of $|E\rangle$ state on the 
    Bloch sphere. (B) Population evolution for the logical 
    states $|E\rangle$ and $|G\rangle$.}
\end{figure}

A few remarks are now in order regarding the different kind of 
excitonic polarization  we have considered so far.
In the second -polarization-based - encoding 
 we need a more laser pulses (and then longer 
time for the application of the gates) respect to the model with 
the first scheme with non-polarized excitons. This makes the set-up slightly more complicated 
but now the logical $1$ and $0$ states corresponds here to energetically
degenerate states with the same orbital wave function structure.
This facts should 
1) make the qubit more robust against {\em pure} dephasing processes
2) set to zero the qubit self-Hamiltonian i.e., the  $\sigma_z$ component
allowing for a simplified gate design and then no recoupling pulse are 
required.

On the other hand it should be be noted that in in the second scheme
both the codewords correspond to {\em unstable} states, indeed excitons 
will eventually recombine through the semiconductor gap by emitting a photon.
On the contrary in the first encoding scheme the logical $0$ corresponds
to the ground state $|G\rangle$ of the crystal, and it is therefore a
stable state. 

Exciton recombination corresponds
in the first scheme    to the amplitude-damping process
$|1\rangle\mapsto|0\rangle.$ One can take care of this kind of environment-induced error
by the both the techniques of quantum error correction \cite{ERR} or error avoiding
\cite{EAC} depending on the spatial symmetry of the damping process. 
Using polarization encoding spontaneous decay gives rise to a leakage
to the computational subspace in the the ground state of the crystal $|G\rangle$
is no-longer a computational codeword. In this case one can resort
to leakage-elimination strategies based of active intervention on the system\cite{leak}.

\section{Simulations}

To test our models
we performed numerical simulations of the quantum gates solving the 
Schroedinger equation.
For the first model (with no polarized excitons) we took $|E\rangle$ 
as starting state and then simulate the evolution when we apply the 
pulse sequences presented. 
In Fig. \ref{fig:fig_gate1} the result of the simulation 
for gate $1$ are shown;
the parameters are chosen in order to obtain a NOT 
gate. In Fig. \ref{fig:fig_gate1} (A) the curve traversed by the 
state in the Bloch space and (B) the population evolutions are presented.
Once decided which gate to apply we can have an estimate of the gate time.
For this NOT gate the laser frequency is not resonant and is constrained 
by the gate choice ($\omega_L = \omega_0 - 2 \Omega$); the time gate is 
fixed by the Rabi frequency of the laser. For realistic laser
parameters ($\Omega^{-1} = 50 fs$) we have : $t_{gate1} = 0.1~ps$.

\begin{figure}[t]
  \includegraphics[height=4.0cm]{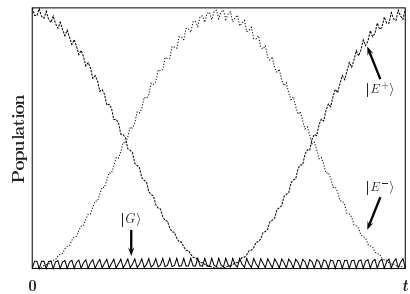}
  \caption{\label{fig:raman} Population evolutions for the 
    three-level system with polarized (logical) excitons 
    $|E^+\rangle$, $|E^-\rangle$ and $|G\rangle$ with lasers with a $\Delta$ 
    detuning. The perturbative parameters is $\Delta / \Omega =10$.}
\end{figure}

In Fig. \ref{fig:fig_gate2}  we show (for gate $2$) the loop in the Bloch 
space (A) the population evolutions  and the phase accumulated 
during the evolution (inset) (B).
The parameters are chosen in order to obtain $\tilde{\gamma}= \pi /4$ 
and the final state is $(1+i)/\sqrt{2}|E\rangle$.
The laser frequency is resonant with the transition ($\omega_L=\omega_0$) 
and with the same Rabi frequency used before we have:
$t_{gate2} = 0.15~ ps$.

In the second model first we have to test the validity of the approximation
used in (\ref{H2});
for this purpose we simulated the evolution of the three-level 
system showed in Fig. \ref{fig:subspaces} and show the result in Fig.
\ref{fig:raman}. We choose 
$\Delta / \Omega =10$  ($\Omega_+ = \Omega_-= \Omega$) and, as we can see, 
this is sufficient to avoid population of $|G\rangle$ state and to have
the standard Rabi oscillations between the logical states.

\begin{figure}[t]
  \includegraphics[height=4.0cm]{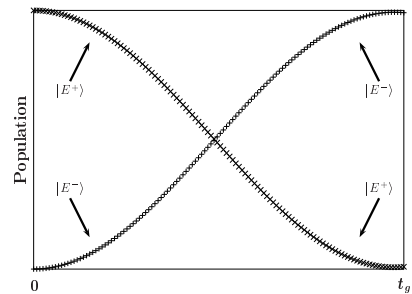}
  \caption{\label{fig:gate3} Populations of logical states 
    for polarized exciton model. The phase accumulated in a single loop is 
    $\gamma=0.0270254$ and we iterate the cycle of $\pi-$pulse $59$ times 
    to obtain a NOT gate.}
\end{figure}

We note that, because of the perturbative request in \ref{H2}, 
the {\it effective} magnetic field ${\bf B}$ has small $x$ and $y$ component,
and then a sequence of two $\pi-$pulse is not sufficient to construct a 
generic superposition of logical qubits. Even if the geometrical phase 
accumulated during the loop is small it is sufficient to iterate the 
procedure to apply the desidered geometrical operator.
Using the same perturbation parameter as in \ref{fig:raman}
we simulate the evolution of $|E^+\rangle$. 
In Figure \ref{fig:gate3} we show the population evolutions of the states 
$|E^+ \rangle - |E^- \rangle$ when they are subjected to a 
$\pi-$pulse sequence in order to obtain a NOT gate. 
Of course the gating time in this situation depends on which gate we want 
to apply and the parameter used the model.

\section{Conclusions}
In summary, we proposed  two approaches to geometric non-adiabatic 
quantum information processing in semiconductor quantum dots. 
In both cases we have been  able to construct a universal set of quantum gates 
using the Aharonov-Anandan phase.
In the first scheme the qubit is realized by the presence or absence of a (ground) state exciton.
A coupling with an external laser field allows for the non-adiabatic realization of 
the geometrical-gates. The dipole-dipole coupling between excitons plays an essential
role in action of the  entangling two-qubit gate.

In the second  approach we encode  information in degenerate states using,
as quantum degree of freedom, the polarization i.e., total spin, of the excitons ($|E^\pm\rangle $).
The logical states are not directly connected but we showed, first how to 
avoid this problems with two-photon (Raman) transition and second how to 
implement in this way a {\it selective phase} gates
(for one and two qubits). 
Numerical  simulations with realistic parameters show that these gates 
can be in principle enacted within the decoherence time. 
The models for non-adiabatic (fast) QIP presented in this paper combine the features of
geometrical gates  with the ultra-fast gate control possible in semiconductor nanostructures;
an experimental verification of these schemes seems under the reach of current technology.

\end{document}